\documentclass[preprint,aps,floats,superscriptaddress]{revtex4}
\usepackage{amsmath}
\usepackage{graphicx}
\usepackage{epsfig}
\begin{document}
\newcommand{\dg}{^{\dagger }}
\newlength{\upit}\upit=0.1truein
\newcommand{\raiser}[1]{\raisebox{\upit}{#1}}
\newlength{\bxwidth}\bxwidth=1.5 truein
\newcommand\frm[1]{\epsfig{file=#1,width=\bxwidth}}
\def\fig#1#2{\includegraphics[height=#1]{#2}}
\def\figx#1#2{\includegraphics[width=#1]{#2}}
\newlength{\figwidth}
\newcommand{\fg}[3]
{
\begin{figure}[ht]
\[
\includegraphics[width=\figwidth]{#1}
\]
\vspace*{-4mm}
\caption{\label{#2}
\small
#3
}
\end{figure}}
\title
{\bf Quenched disorder formulation of the pseudo-gap problem}

\author{A. Posazhennikova}
\email[Email:]{anna@tkm.physik.uni-karlsruhe.de}
\affiliation{
Laboratorium voor Vaste-Stoffysica en Magnetisme,
 Katholieke Universiteit Leuven,
 Celestijnenlaan 200 D, B-3001 Leuven, Belgium}

\author{P. Coleman}
\affiliation{Center for Materials Theory,
Department of Physics and Astronomy, Rutgers University, Piscataway, NJ
08854, USA}

\preprint{version of \today}


\begin{abstract}
The problem of pseudo-gap formation in an electronic system,
induced by the fluctuations of the order parameter is revisited.
We make the observation that a large class of current theories are
theoretically equivalent to averaging the Free energy of the
pseudo-gap system over quenched-disordered distribution of the
order parameter. We examine the cases of both infinite and finite
correlation length, showing how the interplay of pseudo-gap
formation and superconductivity can be treated in this approach.

\end{abstract}
\maketitle

\section{Introduction}

The discovery of a marked suppression in the density of states of
the underdoped cuprate superconductors above the critical
temperature has led to a revival of interest in the idea of a
pseudo-gap.  This concept was originally introduced by Lee,  Rice
and Anderson \cite{LRA} to explain the suppression of electron
density of states associated with order parameter fluctuations
near a charge density wave instability. According to one school of
thought, a similar mechanism may drive the formation of the
pseudo-gap in the underdoped cuprates(see reviews
\cite{Randeria,Timusk,Sadreview}). Many candidate fluctuating
order parameters, such as antiferromagnetism\cite{anti}, charge
density wave\cite{cdw} and pre-formed Cooper pairs have been
proposed \cite{preformedpairs}.

Pseudo-gaps are likely to be a widespread feature of correlated
electron systems lying close to an instability that gaps part of
the Fermi surface, and they have been observed in a wide variety
of nested and low dimensional strongly correlated electron
materials, including one dimensional charge density wave
systems\cite{1dcdw}, vanadium doped Chromium\cite{rosenbaum},
colossal magneto-resistance compounds\cite{cmr} and
strontium-calcium and strontium-barium ruthenates\cite{noh}.

The fluctuation gap model (FGM), based on the idea of the
pseudo-gap emergence due to fluctuating order parameter was
pioneered in \cite{LRA}. Later on it was shown by
Sadovskii\cite{Sadovskii74,Sadovskii79} how the problem of nested
electrons moving in a critically fluctuating order parameter field
could be solved exactly. Interest in this model was revived with
the discovery of the pseudo-gap in cuprates, the physical origin
of which is still controversial. Schmalian et al \cite{Schmalian}
extended the 1D model to higher dimensions and nonzero spin.
Tchernyshyov \cite{Tchernysh} later questioned the exact
solvability of the model for finite correlation lengths and
pointed out a previously unnoticed mistake in the Sadovskii
result. Couple of years ago Millis and Monien\cite{Millis}
performed a careful study of different approximations, explaining
for pseudo-gaps, also by Bartosch and Kopietz \cite{KopietzDOS}
the exact numerical calculation of the density of states in FGM
was carried out. They demonstrated \cite{Millis,KopietzDOS} that
Sadovskii approach is in fact a good approximation and the terms,
omitted in his model produce negligible corrections. In a related
development, Kopietz et al \cite{Kopietz} presented the exactly
solvable model of extended FGM, where fluctuations of both phase
and amplitude of the order parameter have been taken into
account.The important issues of crossover from Gaussian to
non-Gaussian order parameter fluctuations in one-dimensional
Peierls systems have been recently discussed by Monien
\cite{Monien}.

 A key idea of the Lee, Rice and Anderson paper \cite{LRA}is that
order parameter fluctuations can be
regarded as classical degrees of freedom with time-independent
correlation functions.   This basic assumption
threads through most of the subsequent developments during past twenty
years.
This paper revisits this basic idea, making the observation
that a large class of current theories
are theoretically equivalent to averaging the Free energy of the
pseudo-gap system  over a {\sl quenched-disordered} distribution
of the order parameter.  The intuitive equivalence between the slow
critical fluctuations
responsible for pseudo-gap formation and a
quenched random potential was certainly known to several authors in
the past\cite{LRA,DeGennes,Fisher,Chandra,Monthoux}. In this paper, we
vocalize this equivalence and set it in a formal framework.
Using the idea we show how the Sadovskii
model of pseudo-gap formation can be formulated in terms of
a quenched average free energy. To illustrate the utility of this
method, we develop a toy model for the interplay of pseudo-gap
formation  and weak impurity scattering with d-wave superconductivity.
Our results may be of interest to recent experiments on the cuprate
superconductors.


\section{Classical interactions imply quenched disorder}

One of the central assumptions of a large class of pseudo-gap
models, is that the order-parameter fluctuations which scatter
electrons to form the pseudo-gap  are so slow that they can be
considered to be infinitely  retarded.  Such a state of affairs
could come about in a variety of ways. The most pragmatic
point-of-view, is that the characteristic relaxation timescale of
the fluctuations $\tau_{OP} $ is not actually  infinite, but may
be treated as such because it is much greater than the inelastic
scattering rate of the electrons $\tau _{e}$
\[
\tau _{OP}>> \tau _{e}.
\]
The result of this condition, is that electrons perceive the order parameter
as a frozen degree of freedom, that may be treated as a classical
variable.

We now need to consider the interactions between the electrons induced
by these classical order parameter fluctuations.
In a Feynman diagram, the net interaction induced by these fluctuations
has a number of important features (Fig. 1.):
\begin{itemize}
\item It transfers no energy, and is thus infinitely retarded. (Fig. 1 (a))

\item  Polarization bubble insertions into the
effective interaction are absent, since these have already been taken into
account in forming the effective interaction (Fig. 1(b)).

\item
Interactions
between the classical fluctuations may be ignored.
This is  a useful,
assumption, for it leads to non-interacting Gaussian fluctuations and
permits the summation of a large class of diagrams.  Unfortunately,
it
can not really be justified near a critical point,
where non-linear  interactions
between classical fluctuations are almost always relevant.
This is without doubt a major weakness of the
current class of theories, and one we shall return to in our final discussion.
\end{itemize}
\figwidth=0.7\textwidth \fg{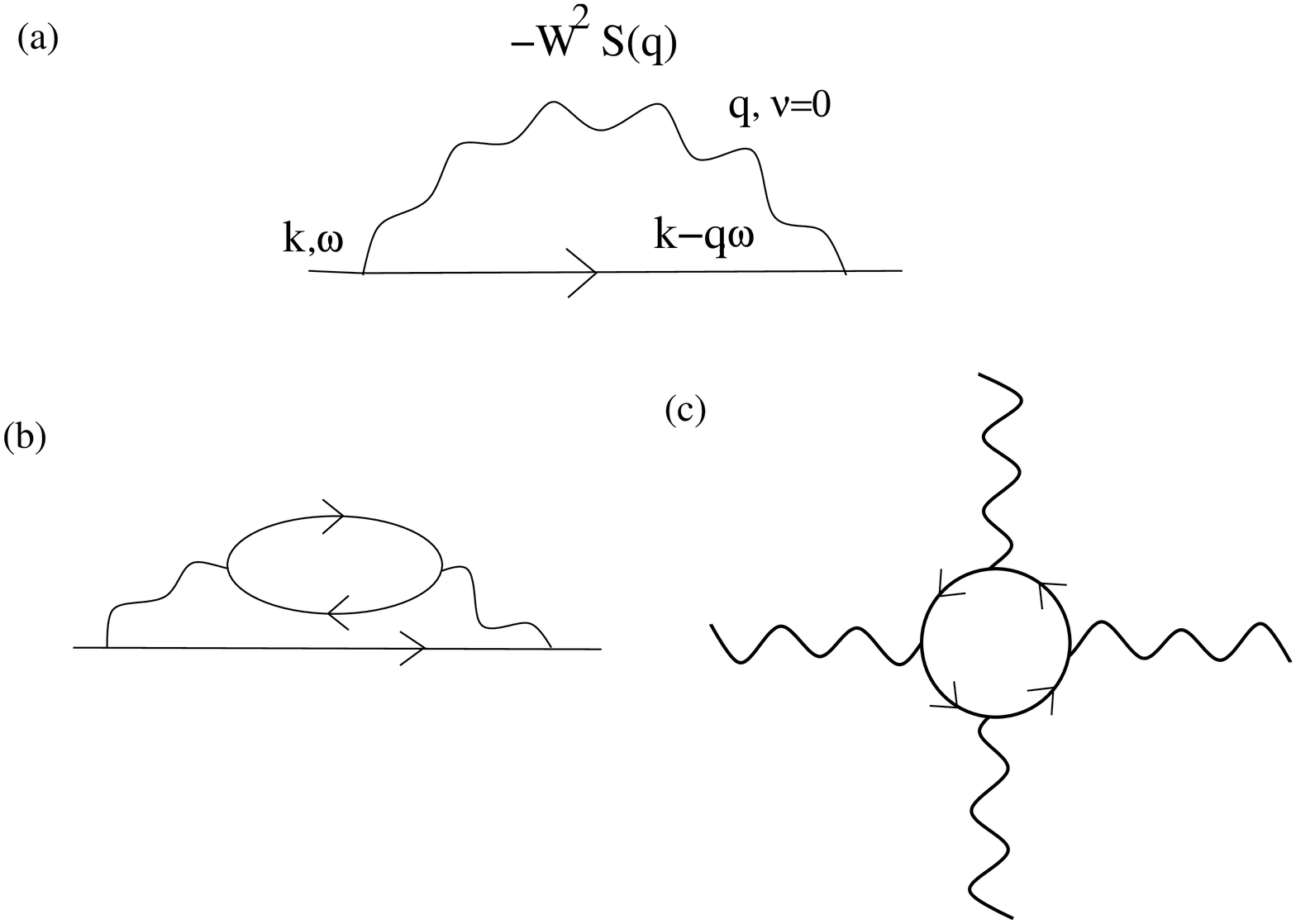}{fig1}{ (a) Classical
fluctuations induce an interaction $V_{eff}= - W^{2}S (\vec{q})$
on a single line. (b) Fermion loop insertions into the interaction
line are neglected which means that the average over the classical
fluctuations is a quenched average. (c) By carrying out a quenched
average, we also eliminate interactions between fluctuations of
the form shown here. } The first feature can be incorporated into
a model by introducing a coupling into the Hamiltonian
\begin{equation}\label{ham1}
{H}[\phi ]= H_{o} + W\sum_{\vec q}\rho _{\vec{q}} \phi
_{-\vec{q}}.
\end{equation}
In this paper we will imagine that $H_{0}$ is a Hamiltonian in
which effects of  interaction have been taken into account by a
mean-field approximation- so that for instance, $H_{0}$ might be a
BCS Hamiltonian with explicit pairing terms. In (\ref{ham1}) $\rho
_{\vec{q}}= \sum_{\vec{k}} c \dg _{\vec{k}-\vec{q}}c_{\vec{k}}$ is
the density of electrons at momentum $\vec{ q}$ and $\phi
_{\vec{q}}$ is a purely classical field representing the order
parameter fluctuations.  This quantity is to be averaged over the
distribution
\[
P[\phi ] = exp\left[- \beta \sum
\frac{\vert \phi _{\vec{q}}\vert ^{2} }{ 2 S(\vec{q})}
\right]
\]
and it thus gives rise to an effective, and infinitely retarded
(attractive)  interaction of the form
\[
V_{eff} (\vec{q})= -W^{2} S (\vec{q})
\]
At first sight, it might appear sufficient average the
partition
function over $P[\phi] $, i.e
\begin{eqnarray}\label{}
Z&=& \int P[\phi ] Z[\phi ]\cr
Z[\phi ]&=& {\rm Tr}\left[e^{-\beta H[\phi ]} \right].
\end{eqnarray}
This is an \underline{annealed} average.
However, such a scheme allows fermion loop renormalizations and
non-linear interactions to develop within the
interaction lines as illustrated in Fig. 1 (b). These
terms  are to be dropped, under the assumptions mentioned above.
To ensure this,
we must carry
out a \underline{quenched average} over the classical field,
which means that we average the Free energy, rather than the partition
function
\begin{eqnarray}\label{}
F = -T \int P[\phi ] \log Z[\phi ]
\end{eqnarray}
This  procedure is implicit to the approach of
Sadovskii, and subsequent developments, by the selection of Feynman
diagrams containing a single Fermion line.
One way to implement the quenched average is to use the
$N\rightarrow 0$
trick,
\begin{eqnarray}\label{quenchedlimit}
F = -T\lim_{N\rightarrow 0}\int P[\phi ] \frac{1}{N}\left(Z[\phi
]^{N}-1\right).\end{eqnarray}
The term containing $Z[\phi ]^{N}$ can be considered as the partition function
of N identical replicas in the frozen classical field $\phi $.
Formally, this is accomplished by introducing $N$ replicas of the
fermion fields, labeled by the index $\lambda \in [1,N]$. The
partition function is then given by
\[
Z^{N}[\phi ]=\hbox{Tr}[e^{-\beta H^{( N)}[\phi ]}]
\]
where
\begin{equation}\label{}
H^{( N)}[\phi ]= \sum_{\lambda =1,N}\left(H^{\lambda }_{o} + W\sum_{\vec
q}\rho^{\lambda } _{\vec{q}} \phi
_{-\vec{q}}
 \right).
\end{equation}
is the Hamiltonian of $N$ identical replicas coupled to a single
classical field.
Diagrammatic contributions to $Z[\phi ]^{N}$
containing $n$ fermion loops, scale as $O ( N^{n})$, so that in the
$N\rightarrow 0$ limit, the only terms surviving in the diagrammatic
expansion of the Free energy, are those with a single fermion loop, as follows
 \bxwidth=0.2\textwidth \upit=-0.28truein
\begin{eqnarray}\label{}
\Delta F &=& -T\lim_{N\rightarrow 0}\left[\frac{1}{N}\left( \overbrace
{\raiser{\frm{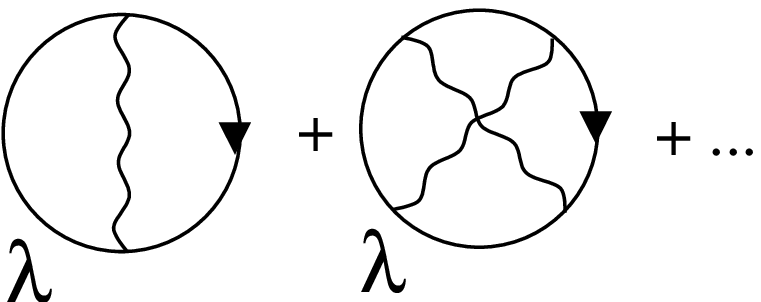}}}^{O (N)}\quad
 \bxwidth=0.3\textwidth \upit=-0.38truein
\overbrace {\raiser{\frm{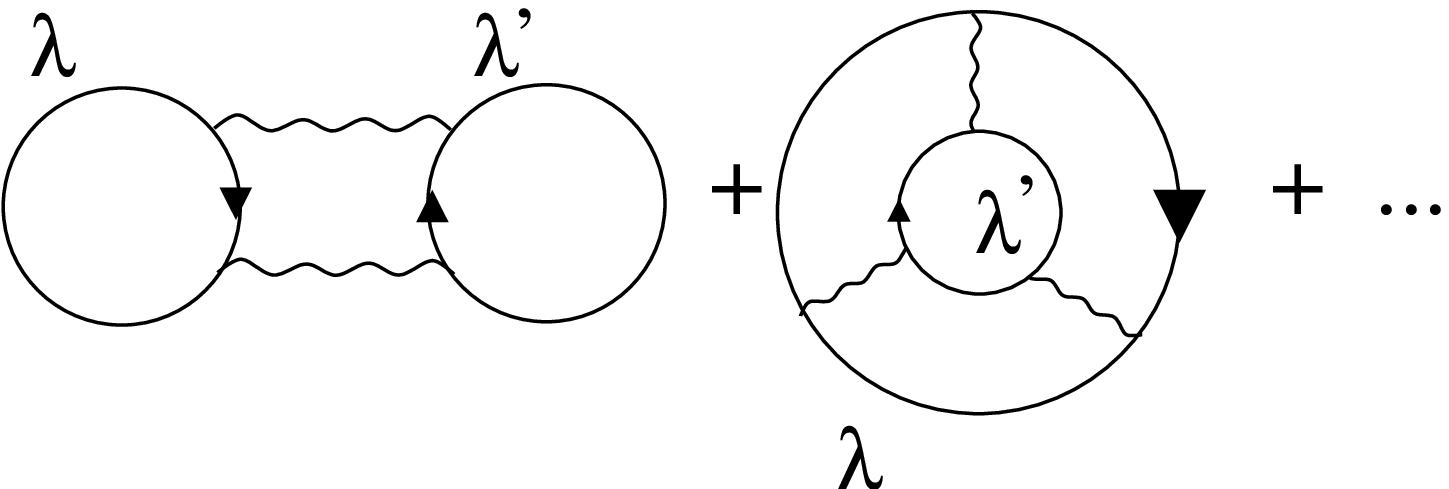}}}^{O (N^{2})} \right) \right]
,
\end{eqnarray}
where $\Delta F=F_{I}-F_{0}$ is the change in the Free energy due  to
turning on the interactions,
Thus the
only terms which survive the $N\rightarrow 0$ limit are those with
a single fermion line or loop\cite{Edwards}. In this way, all RPA
renormalizations {\sl and} non-linear interactions between the
fluctuation fields are eliminated, as follows,
\begin{equation}\label{}
\bxwidth=0.5 \textwidth \upit=-0.2truein \Delta F= -
T\left[\raiser{\frm{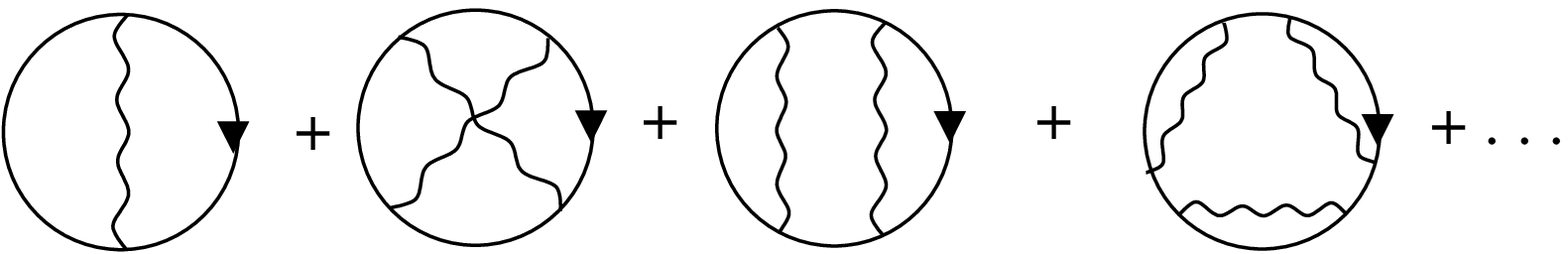}} \right],
\end{equation}
where the replica index has been eliminated from the fermion lines.
The advantage of this approach is that we can now start to formulate a theory of
the pseudo-gap in the
language of an effective Free energy.  This becomes particularly
useful when considering the case of a superconductor with a
pseudo-gap.

\section{Effective action in the pseudo-gap model with infinite correlation
length}

To illustrate the above approach, we consider a two-dimensional
electronic system with  a nested Fermi surface spanned by the
commensurate wavevector $\vec{Q}= (\pi ,\pi )$ as illustrated in
Fig. 2. Fluctuations of the underlying order parameter are
governed by a correlation function of the form
\begin{equation}
S(\vec{q})=\frac{1}{\pi^2}\frac{\xi^{-1}}{(q_x-Q_x)^2+\xi^{-2}}
\frac{\xi^{-1}}{(q_y-Q_y)^2+\xi^{-2}} \label{fluct}
\end{equation}
fluctuations couple the nested Fermi surfaces via the effective
interaction $V_{eff}=-W^2S(\vec{q})$. \figwidth=0.5\textwidth
\fg{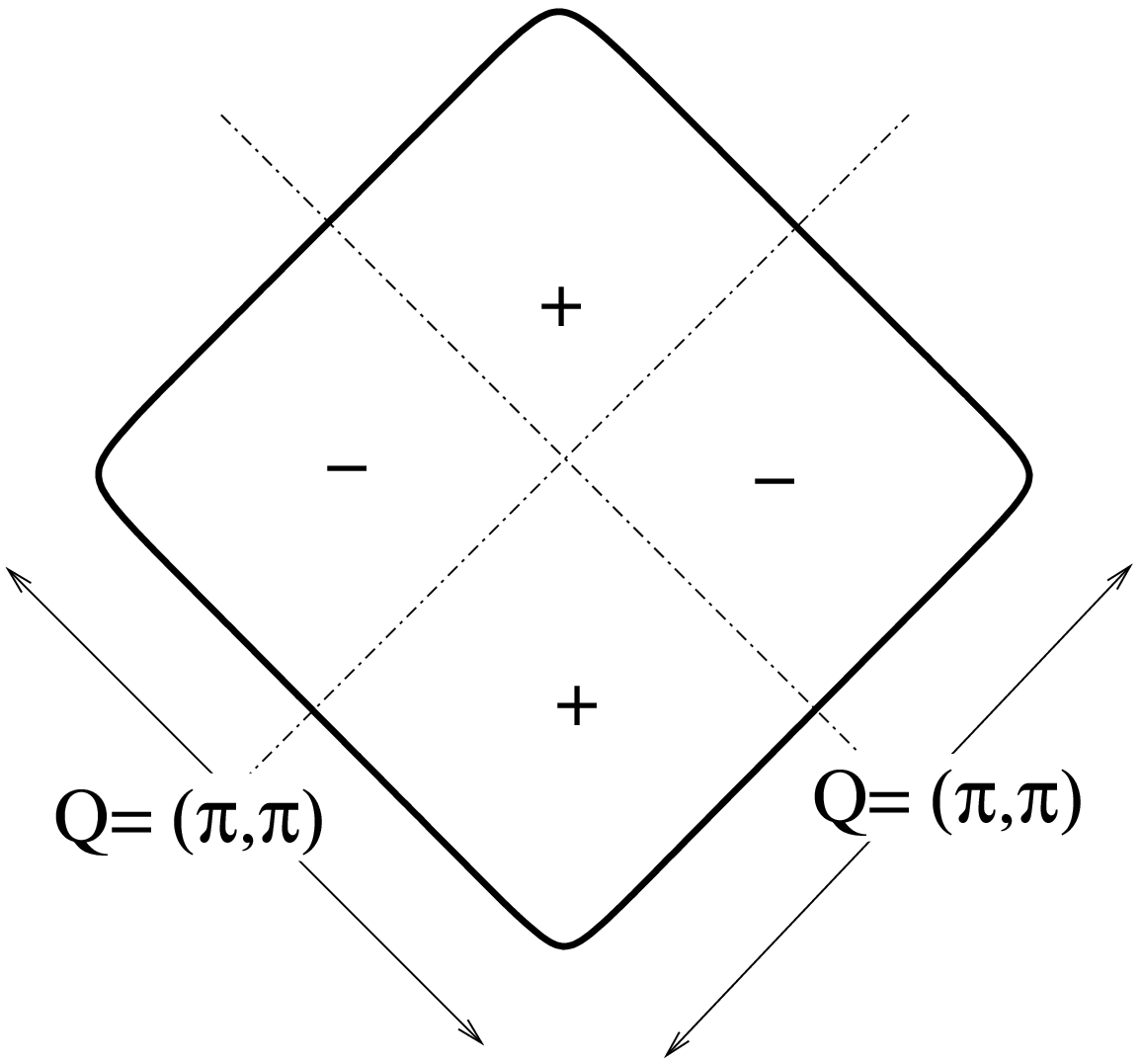}{fig2}{Nested Fermi surface with commensurate
fluctuations considered in the toy model of this paper.} It was
shown \cite{PosSad} that in the limit $\xi \rightarrow \infty$ the
model under consideration can be solved exactly and the effective
interaction takes the oversimplified form
\begin{equation}
V_{eff} (\vec{q}) =- W^2\{\delta(q_x-\pi )\delta ( q_y-\pi )\}.
\label{WW}
\end{equation}
In this paper, we will compliment the earlier work\cite{PosSad} by
showing how the diagrams for the Free energy may be completely summed
in this limit.
The class of models we shall discuss take the form
\begin{equation}\label{}
H= H_{0}+ H_{I}
\end{equation}
where $H_{0}$ is a Hamiltonian which does not scatter between the
nested Fermi surfaces and $H_{I}$ contains the coupling to the density
fluctuation modes.
We begin by considering the simple model where
\begin{eqnarray}\label{mod1}
H_{0}&=& \sum_{\vec{k}}\epsilon _{\vec{k}} c\dg _{\vec{k}\sigma }
c _{\vec{k}\sigma }\cr 
 H_{I}&=&  W \sum_{\vec{k},\vec{q}}c\dg
_{\vec{k}+\vec{q}} c_{\vec{k}}\phi _{\vec{q}},\end{eqnarray} where
$\phi _{\vec{q}}$ is a classical field described by quenched
average over the probability distribution $P[\phi ]$ given above.
If we denote the effective action by a wavy line, then the Feynman
diagrams for the expansion of the Free energy are
\bxwidth=0.7\textwidth \upit=-0.4truein
\begin{eqnarray}\label{}
\cr
\raiser{\frm{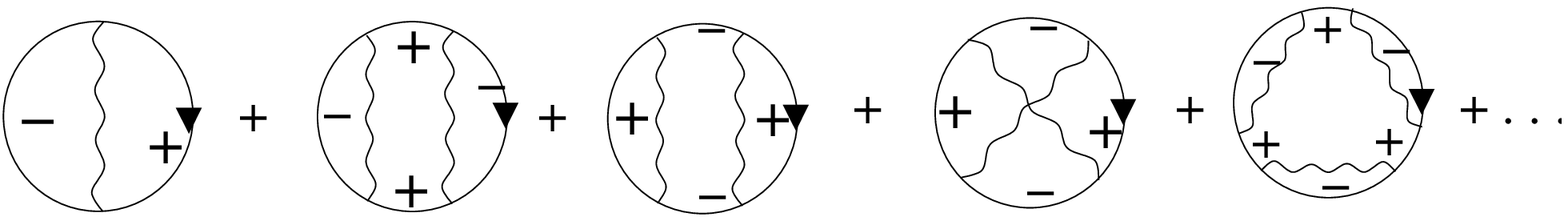}}\cr.
\end{eqnarray}
Here we have implicitly taken the $N\rightarrow 0$ limit of
(\ref{quenchedlimit}), thereby eliminating all
polarization renormalizations of the interaction lines. We have
used the pseudo-momentum labels $``-"$ and $``+"$ to denote
electrons lying on the ``left'' or ``right''-hand side of the
nested Fermi surface. An essential feature of our derivation is
that the pseudo-momentum alternates between scattering events.
(For incommensurate fluctuations, this requirement eliminates some
of the diagrams such as the fourth diagram shown above). In the
limit of an infinite correlation length, each interaction line can
effectively be replaced by a static modulated scattering potential
which scatters at the nesting wavevectors. Consider the simplest
$W^{2}$ diagram. In the limit of infinite correlation length, the
interaction can be replaced by a delta function in momentum, so
that \bxwidth=0.3\textwidth \upit=-0.4truein
\begin{eqnarray}\label{}
\cr
\raiser{\frm{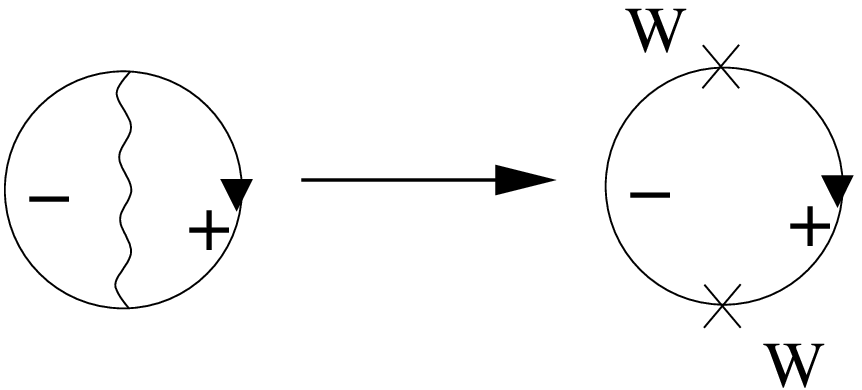}}\cr
\end{eqnarray}
where we have represented the vertex of the
static scattering potential by a ``cross''.
When we come to consider higher order diagrams, each term
of a given order gives rise to the same contribution, however,
we must be careful to take into account symmetry factors.
Let the number of closed diagrams of order $W^{2n}$ be $\alpha _{n}$, then
\bxwidth=0.17\textwidth
\upit=-0.4truein
\begin{eqnarray}\label{}
\cr
\sum
\left\{\raiser{\frm{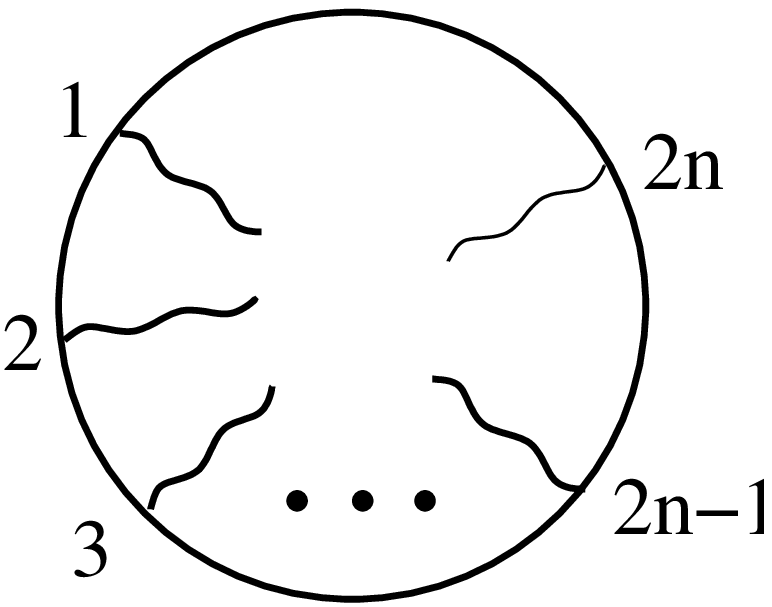}}
 \right\}
\longrightarrow \alpha _{n}
\bxwidth=0.22\textwidth
\upit=-0.4truein
\left\{
\raiser{\frm{{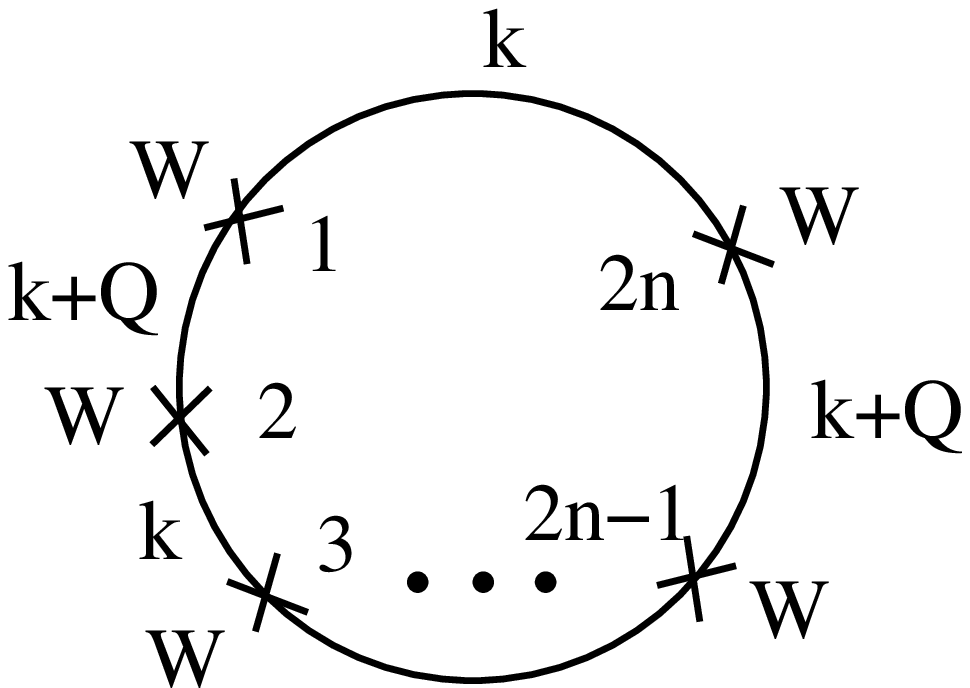}}} \right\}\cr
\end{eqnarray}
Around the perimeter of each diagram of this order, there are $n$
propagators of both ``$-$'' and ``$+$'' pseudo-momentum. If we
differentiate the sum of all $2n$th order diagrams with respect to
$G_{\vec{k}}$, we  generate the $2n$th order self-energy diagram,
as follows \bxwidth=0.17\textwidth \upit=-0.4truein
\begin{eqnarray}\label{}
\cr
\frac{\delta }{\delta G_{\vec{k}}}
\sum
\left\{ \raiser{\frm{fig4a.eps}}\right\}
= \sum\left\{
\bxwidth=0.21\textwidth
\upit=-0.4truein
\raiser{\frm{{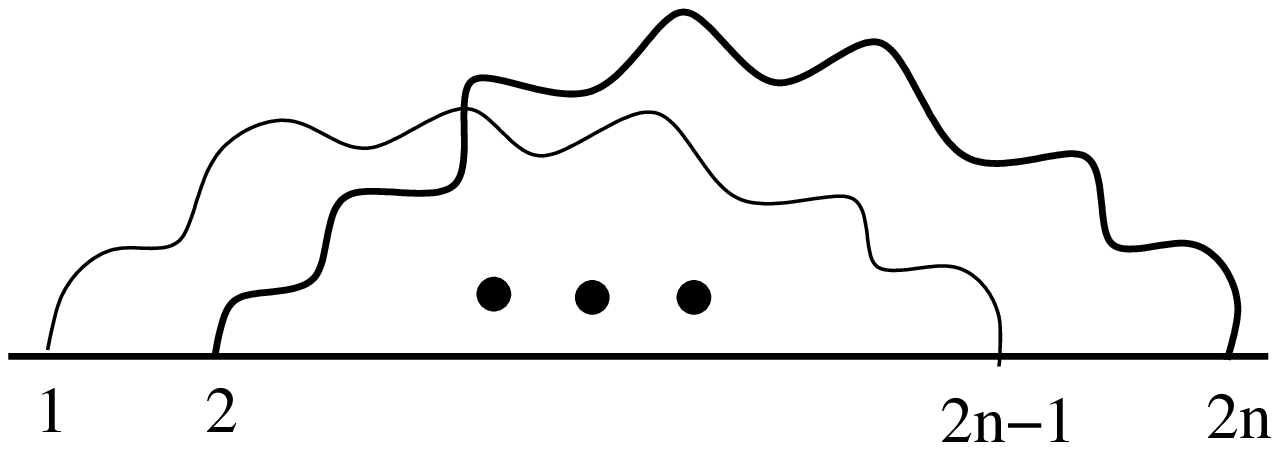}}} \right\}\cr
\end{eqnarray}
But in the limit of infinite correlation length, this is equal to
\bxwidth=0.17\textwidth
\upit=-0.4truein
\begin{eqnarray}\label{}
\cr
\alpha _{n}\frac{\delta }{\delta G_{\vec{k}}}
\bxwidth=0.21\textwidth
\upit=-0.4truein
\left[\raiser{\frm{fig4b.eps}} \right]
= \alpha _{n}\left\{
\raiser{\frm{{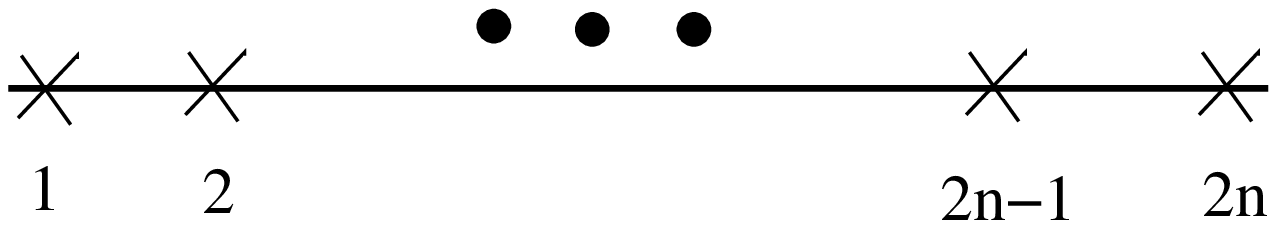}}} \right\}\cr
\end{eqnarray}
Now the number of $2n$ th order self energy diagrams is given by
the number of ways of connecting n interaction propagators to
2n interaction vertices so that
\bxwidth=0.2\textwidth
\upit=-0.4truein
\begin{eqnarray}\label{}
\cr
\sum
\left\{\raiser{\frm{fig4c.eps}} \right\}
= (2n-1)!!\left\{
\raiser{\frm{{fig4d.eps}}} \right\}\cr
\end{eqnarray}
enabling us to identify $\alpha _{n}= (2n-1)!!$.
Using the relation
\[
(2n-1)!!= \frac{2^{n}}{\sqrt{\pi }} (n-\frac{1}{2})!
\]
where  we denote $(x-\frac{1}{2})!\equiv \Gamma (x+\frac{1}{2})$, we
can absorb the factor  $(2n-1)!!$ into
an integral over a Gaussian {\sl distribution}
of gap sizes as follows
\begin{eqnarray}\label{distribution}
W^{2n}(2n-1)!!&=& \frac{1}{\sqrt{2\pi }}\int_{-\infty }^{\infty }  {d\zeta
}{}e^{-\zeta^{2} /2}
({\zeta } W)^{2n},
\end{eqnarray}
in other words
\bxwidth=0.17\textwidth
\upit=-0.35truein
\begin{eqnarray}\label{}
\cr
\sum
\raiser{\frm{fig4a.eps}}
=
\frac{1}{\sqrt{2\pi }}\int_{-\infty }^{\infty }  {d\zeta
}{}e^{-\zeta^{2} /2}
\left\{
\bxwidth=0.22\textwidth
\upit=-0.35truein
\raiser{\frm{{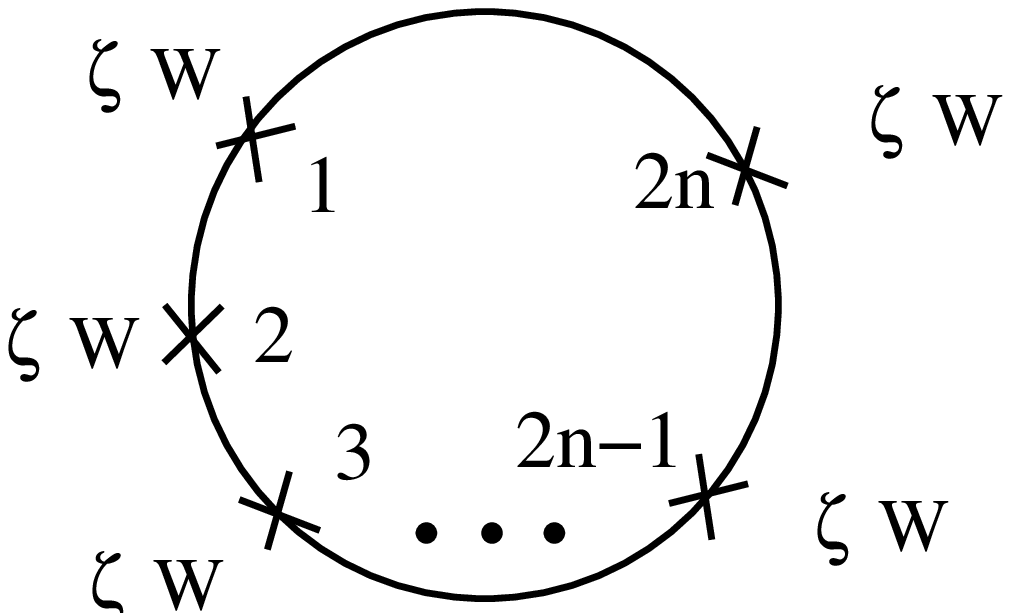}}} \right\}\cr
\end{eqnarray}
Remarkably- all that is left of the average over the classical fields
is a residual distribution over pseudo-gap sizes.
In other words, in the limit of an infinite correlation length,
the quenched average over the distribution of scattering
potentials is replaced by a single average over a
Gaussian distribution
of scattering strengths. The quenched averaged Free energy is then
given by
\bxwidth=0.22\textwidth
\upit=-0.35truein
\begin{eqnarray}\label{}
&&\cr
F&=& F_{o}+
\frac{1}{\sqrt{2\pi }}\int_{-\infty }^{\infty }  {d\zeta
}{}e^{-\zeta^{2} /2}
\sum_{n}\left\{
\bxwidth=0.22\textwidth
\upit=-0.35truein
\raiser{\frm{{fig4e.eps}}} \right\}\\
\nonumber
\end{eqnarray}
where $F_{o}= -T\sum_{\vec{k},\ {n}}\ln [-G_{o}^{-1} (\vec{k},i\omega_{n})]$
is the Free energy of the ``non-interacting'' system.  Combining
both terms, we then obtain
\begin{eqnarray}\label{}
F &=&  \frac{1}{\sqrt{2\pi }}\int_{-\infty }^{\infty }  {d\zeta
}{}e^{-\zeta^{2} /2}   F[\zeta ,W]
\end{eqnarray}
where
\begin{eqnarray}\label{}
F[\zeta ,W]=-T\sum_{n, \ \vec{ k} \in \frac{1}{2}\hbox{\tiny BZ}}
{\rm  Tr}\ln
\begin{pmatrix}
-G_{o}^{-1} (\vec{k}-\vec{Q}/2,i\omega _{n})
& \zeta W\cr
\zeta W& -G_{o}^{-1} (\vec{k}+\vec{Q}/2,i\omega _{n})
\end{pmatrix}
\end{eqnarray}
is the Free energy for a gap size $\zeta W$.
The most important feature of our derivation above, is the conservation
of the pseudo-momentum associated with each nested Fermi surface.
This  derivation can be
easily generalized to include pairing (which conserves momentum
) and a restricted class of disorder scattering which scatters
between states on the same nested side of the Fermi surface.
We now examine these two cases in detail

\subsection{Pairing in the presence of a pseudo-gap}

To consider pairing,
let us now take
\begin{equation}\label{}
H_{0}= \sum_{\vec{k}}\epsilon _{\vec{k}} c\dg _{\vec{k}\sigma }
c _{\vec{k}\sigma }
+ \sum_{\vec{k}}\left[ \Delta \gamma_{\vec{k}}c\dg
_{\vec{k}\uparrow}c\dg _{-\vec{k}\downarrow }+\hbox{H.c.}\right]
+\frac{\vert \Delta\vert ^{2} }{g}.
\end{equation}
where $\gamma_{\vec{k}}= \cos(k_{x})-\cos(k_{y})$ is a d-wave order
parameter  whose nodes bisect the nested Fermi surfaces.
We have included an additional term $\vert \Delta \vert ^{2}/g$
to take account of the BCS decoupling of the pairing interaction.
To make the model more interesting, we shall consider a
pseudo-gap with d-wave symmetry,
\[
H_{I}= W  \sum_{\vec{k},\vec{q}}\gamma _{\vec{k}}c\dg
_{\vec{k}+\vec{q}} c_{\vec{k}}\phi _{\vec{q}},
\]
where $\gamma_{\vec{k}}= \cos (k_{x})- \cos (k_{y})$ as before.
Since the pairing does not change the  momentum of particles, the
reasoning for a normal state is quickly generalized by using the appropriate
Nambu electron propagators.
Let us first derive the free energy expression for this case. The
mean -field Hamiltonian can be expressed in the following general
form
\begin{equation}\label{hamiltonian}
H=\sum_{k \in \frac{1}{2}Bz} \Psi_{k}^+ {\cal H}
\Psi_{k}+\frac{|\Delta|^2}{g},
\end{equation}
where
\begin{equation}
\Psi_{k}=\left(
\begin{array}{c}
  c_{k+Q/2\uparrow} \\
  c_{-k-Q/2\downarrow}^+ \\
  c_{k-Q/2\uparrow} \\
  c_{-k+Q/2\downarrow}^+
\end{array} \right),
\end{equation}
and
\newcommand{\sizer}{}
\begin{equation}\label{calh}
{\cal H}=\left(
   \begin{array}{cccc}
\sizer   \epsilon_{k+Q/2} & \sizer \Delta_{k+Q/2} & \sizer {W_k} &
\sizer {0} \\
  {\Delta_{k+Q/2}} & \sizer {-(\epsilon_{k+Q/2})} & \sizer {0} &
\sizer {-W_{-k}} \\
  {W_k} & \sizer {0} & \sizer {\epsilon_{k-Q/2}} & \sizer
{\Delta_{k-Q/2}} \\
  {0} & \sizer {-W_{-k}} & \sizer {\Delta_{k-Q/2}} & \sizer {-
\epsilon_{-(k-Q/2)}}
\end{array}\right),
\end{equation}
where
$W_{\kappa}=\sin{\kappa_x}-\sin{\kappa_y}$. Now since
$\epsilon_{\kappa-Q/2}=\epsilon_{-(\kappa+Q/2)}$,
$\Delta_{\kappa-Q/2}=-\Delta_{\kappa+Q/2}$ and
$W_{\kappa}=-W_{-\kappa }$,
we can write
\begin{equation}\label{ham}
{\cal H}=\left(
\begin{array}{cccc}
\sizer {\epsilon_{\kappa+Q/2}} &\sizer {\Delta_{\kappa+Q/2}} &\sizer
{W_{\kappa}} &\sizer {0} \\ {\Delta_{\kappa+Q/2}} &\sizer {-
\epsilon_{\kappa+Q/2}} &\sizer
{0} &\sizer {+W_{\kappa}} \\
\sizer {W_{\kappa}} &\sizer {0} &\sizer {-\epsilon_{\kappa+Q/2}} &\sizer {-
\Delta_{\kappa+Q/2}} \\{0} &\sizer {W_{\kappa}} &\sizer {-\Delta_{\kappa+Q/2}}
&\sizer {\epsilon_{\kappa+Q/2}}
\end{array}
\right)
\end{equation}
so that
\[
{\rm  Tr} \ln[-{\cal G}^{-1} (i\omega _{n})]= \ln \det [{{\cal
H}-i\omega _{n}}]
=\ln[\omega_n^2+\epsilon_{k+Q/2}^2+\Delta_{k+Q/2}^2+W_{k}^{2} ].
\]
so that the
the expression for the quenched-averaged
Free energy takes the form
\begin{equation}\label{free}
F=\frac{1}{\sqrt{2\pi }}\int_{-\infty }^{\infty}d\zeta  e^{-\zeta
^{2}/2} \left\{-2T\sum_{n, \ k\in \frac{1}{2}\hbox{\tiny
BZ}}\ln[\omega_n^2+\epsilon_{k+Q/2}^2+\Delta_{k+Q/2}^2+ (\zeta
W_k)^{2} ]+\frac{\Delta^2}{g}\right\}.
\end{equation}

The gap-equation is provided by the condition that the
derivative of the free energy is equal to zero,
$\frac{\partial F}{\partial \Delta}=0$, i.e
\begin{equation}\label{gap}
\frac{1}{g}+\frac{1}{\sqrt{2\pi }}\int_{-\infty }^{\infty}d\zeta
e^{-\zeta ^{2}/2} \left\{ -2T\sum_{n, \ k\in
\frac{1}{2}\hbox{\tiny BZ}} \frac{\gamma_{k}^{2}}
{\omega_n^2+\epsilon_{k}^2+\Delta_{k}^2+\zeta^{2}W_{k-Q/2} ^{2}}
 \right\}
=0.
\end{equation}
At the critical temperature $T_c$, $\Delta=0$ and we have the
transition temperature equation
\begin{equation}\label{tc}
\frac{1}{g}=\frac{1}{\sqrt{2\pi }}\int_{-\infty }^{\infty}d\zeta
e^{-\zeta ^{2}/2} \left\{ 2T\sum_{n, \ k\in \frac{1}{2}\hbox{\tiny
BZ}} \frac{\gamma_{k}^{2}}
{\omega_n^2+\epsilon_{k}^2+\zeta^{2}W_{k-Q/2} ^{2}}
 \right\}.
\end{equation}
In this way, we have successfully eluded the diagrammatic derivation of the
two-particle Green's function\cite{PosSad}
and derived the main result in a
compact and more general way.

\subsection{Impurity effect}

As a further application of these methods, let us now introduce
disorder into the paired pseudo-gap model of the last section. The
interplay of disorder and pseudo-gap formation on pairing is of
particular interest in the context of underdoped cuprate
superconductors. In a recent experimental paper, Tallon et al.
\cite{Tallon} have shown that the superconducting transition
temperature of $La-241$ and $(Y,Ca)-123$ is much more rapidly
suppressed by disorder in underdoped compounds, where a pseudo-gap
is present. Tallon et al explained their results in terms of a
phenomenological model involving unitary scattering taking place.
We now show how the essential features of their discussion can
also be obtained by treating the co-existence of pseudo-gap
formation and weak non-magnetic impurity scattering.

In the presence of impurity scattering we have to include the
additional terms in our Hamiltonian of the form
\[
H'= \sum_{j,k k '\sigma }e^{-i (k -k ')\cdot \vec{ R}_{j}} V_{k k
'}c\dg _{k \sigma} c_{k '\sigma}
\]
In order to apply the methods of the last section, we make the
important assumption that impurity potential $V_{k k'}$ does
not
scatter between different sides of the Fermi surface.
Such scattering can still
transfer electrons across the node, and thus remains severely
pair-breaking. With this simplified assumption,  the impurity scattering
does not renormalize the strength of the pseudo-gap potential.

We can write now
\newcommand{\avge}
{\frac{1}{\sqrt{2\pi }}\int_{-\infty }^{\infty}d\zeta
e^{-\zeta ^{2}/2}}
\begin{equation}\label{}
F=\avge F[\zeta ,V_{k k'}].
\end{equation}
In this oversimplified model the effect of impurities
reduces just to replacing of thin lines to there ``thick",
impurity dressed equivalents in the graphic representation for the
two-particle Green's function in our model.

Using this result, we can generalize equation (\ref{tc} ) to obtain
\begin{equation}\label{tcequation}
\frac{1}{g}=T_c\sum_n \int_{-\infty }^{\infty} \frac{d\zeta
}{\sqrt{2\pi }} e^{-\frac{\zeta ^{2}}{2}}
\int_{0}^{\infty}\frac{d^2k}{(2\pi)^2}\frac{\gamma
^2(\phi)}{\tilde \omega_n^2+\epsilon_k^2+\zeta ^{2}W^2(\phi)},
\end{equation}
where $\tilde\omega_{n}$ is the usual renormalized Matsubara frequency
\begin{equation}\label{omega}
\tilde\omega_{n}=\omega_{n}+\Gamma sign \omega_{n}
\end{equation}
\begin{equation}\label{gamma}
\Gamma=\pi\rho N(0)V_{imp}^2,
\end{equation}
and $\rho$ is the impurity concentration. 

Notice 
that the
momentum dependence of $W_{k-Q/2}\propto \gamma_{k}$, so that in
equation (\ref{tcequation}), 
the functions $\gamma (\phi )$ and $W (\phi )$ share the same angular dependence around the
Fermi surface.
For simplicity we assume the $d$-wave symmetry of the pseudogap as
well as of the superconducting gap function
\begin{equation}\label{pg}
W(\phi)=W \gamma(\phi),
\end{equation}
where $\gamma (\phi)=\sqrt{2}\cos(2\phi)$ is the angular
dependence of the gap around the Fermi surface \footnote{Another
way to introduce the pseudogap is to assume the appearance of
so-called ``hot" regions on the Fermi surface with non-zero
pseudogap \cite{PosSad}}.

After integrating over energy and carrying out the Matsubara sum,
(Appendix A) we obtain the final equation for the critical
temperature:
\begin{equation}\label{finaltc}
\ln\frac{T_c}{T_{c0}}=\int_0^{2\pi}\frac{d\phi}{2\pi}\gamma^2(\phi)
\int_{-\infty}^{\infty}dx \Phi(x) \left\{
\Psi\left(\frac{1}{2}\right)-\Psi\left(\frac{1}{2}+\frac{\Gamma+ix
W(\phi)}{2\pi T_c}\right)\right\},
\end{equation}
where the normalized
distribution function $\Phi (x)$ can be written in terms of the
modified Bessel function $K_{0} (x)$ as
\[
\Phi (x)=\frac{1}{\sqrt{2\pi ^{3}}}e^{-x^{2}/4}K_{0} \left(\frac{x^{2}}{4}
\right)
.
\]
The relative factor of `` $ i $'' between the impurity and pseudogap
term in this expression reflect the different mechanism by which
impurity scattering and the pseudogap supress superconductivity. 
On the one hand, impurity scattering produces pair breaking, by
scattering particles across the nodes of the order parameter. By
contrast, the pseudogap suppresses superconductivity by removing
states from 
around the Fermi energy.

\begin{figure}[!ht]
\begin{center}
\includegraphics[width=0.7\textwidth]{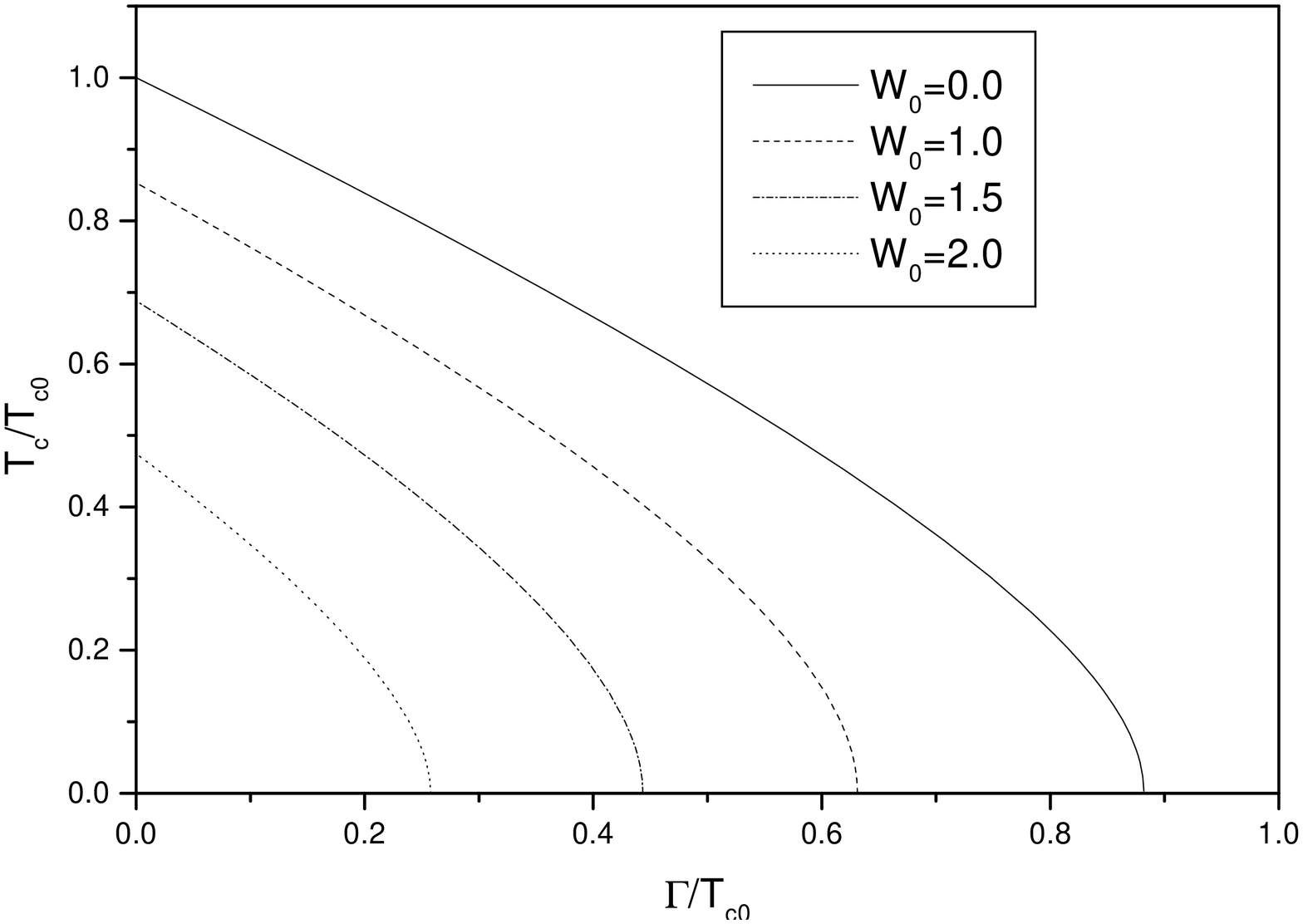}
\includegraphics[width=0.7\textwidth]{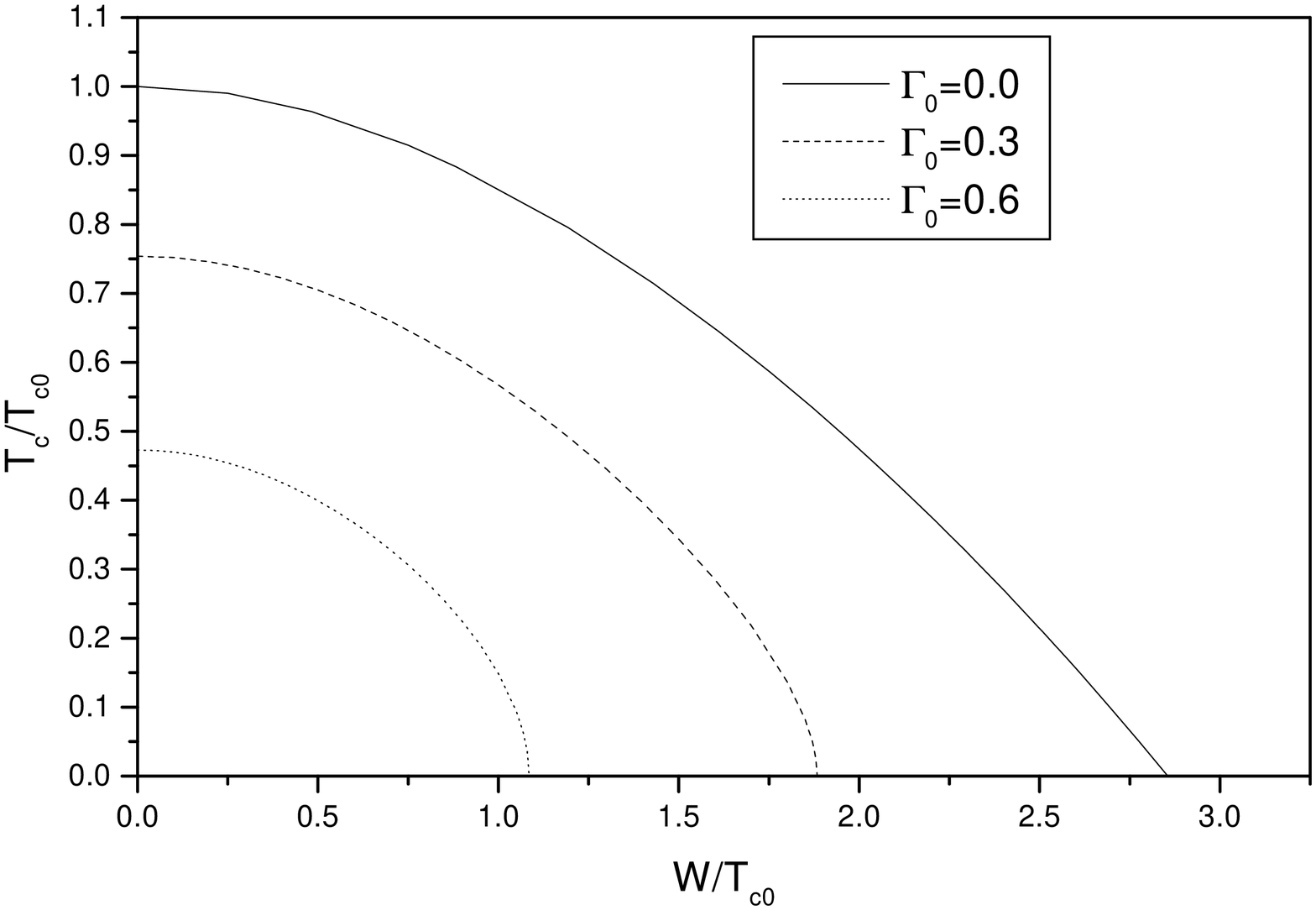}
\end{center}
\caption{(a) Critical temperature of a superconductor, normalized
by the critical temperature of a pure superconductor without
pseudogap $T_{c0}$, vs impurity scattering $\Gamma$ 
in the presence of the pseudogap ($W$ is an effective pseudogap
width). (b) Critical temperature of a superconductor vs the
effective width of the pseudogap for different values of impurity
scattering rate $\Gamma$(both $W$ and $\Gamma$ are normalized by
$T_{c0}$ ). These curves were all obtained from a numerical solution
of equation (\ref{finaltc}).} \label{tcgam}
\end{figure}

 The theoretical results for the critical temperature
dependence on the impurity scattering and the pseudogap, obtained
by a numerical solution of Eq.(\ref{finaltc}), can be presented in
different ways. Fig. 3a, for example, shows the variation of the
transition temperature with impurity scattering rate for a set of
different values of the pseudogap. In the limit where the
pseudogap vanishes, Eq.(\ref{finaltc}) naturally reverts to the
standard Abrikosov-Gorkov mean-field theory. Fig. 3b shows the
dependence of the critical temperature on the pseudogap width for
different ratios $\Gamma_0\equiv \Gamma/T_{c0}$. These figures
illustrate how both the pseudogap and impurity scattering suppress
the superconductivity.

 To illustrate
this approach we now use it to develop a phenomenology for the
cuprate phase diagram (Fig.4). We consider a model, in which the
bare superconducting transition temperature is a linear function
of doping $x$, vanishing at $x=0.3$ and the pseudogap is a rapidly
reducing function of $x$, which vanishes at $x=0.19$. The effect
of the pseudogap is to suppress superconductivity at low doping,
leading to a maximum in the transition temperature in the vicinity
of where the pseudogap goes to zero. When disorder is introduced
the superconducting region is pushed to higher doping.

\begin{figure}[!ht]
\begin{center}
\includegraphics[width=0.5\textwidth]{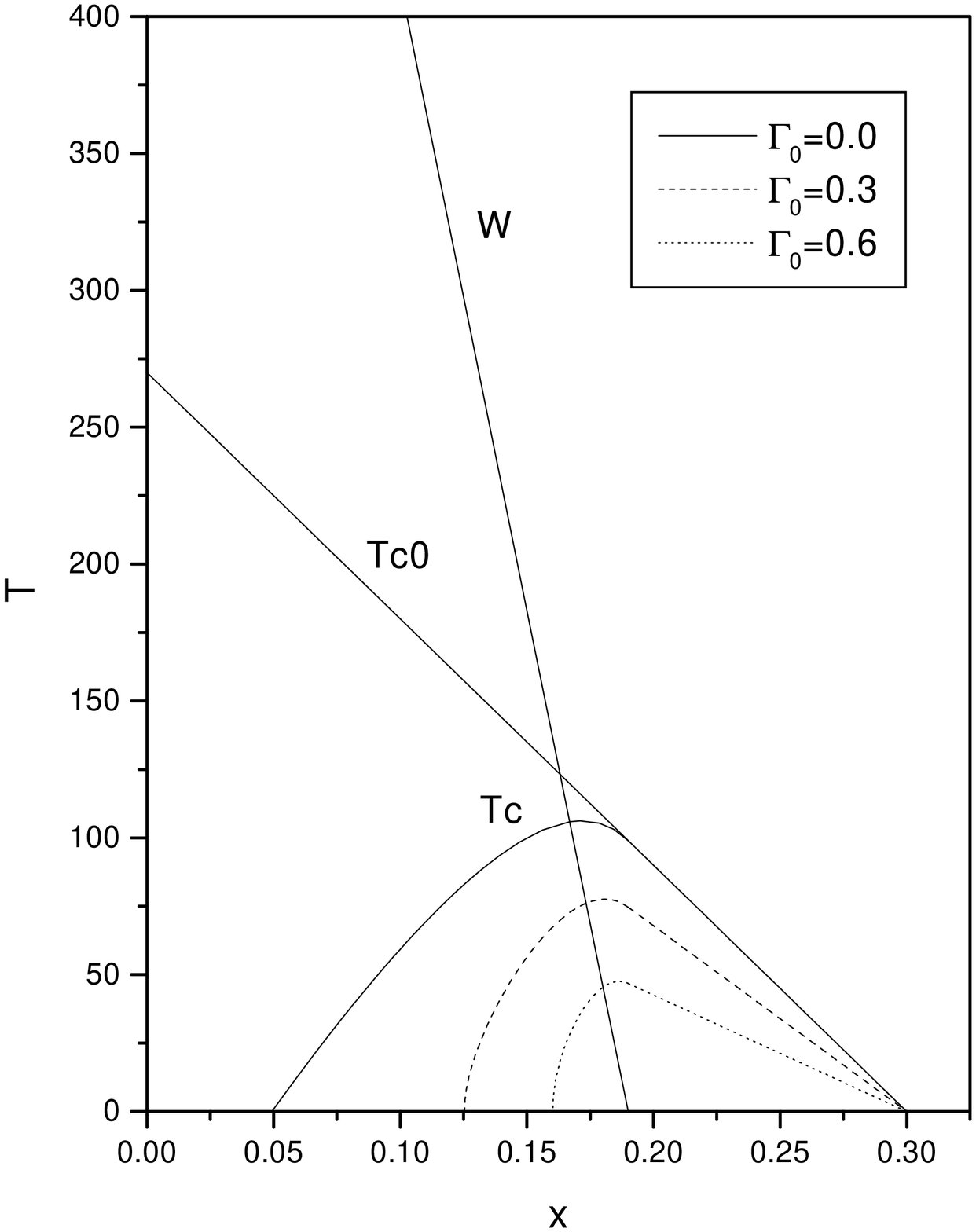}
\end{center}
\caption{Toy phase diagram of cuprates. Here $W$ is the effective
energy scale of the pseudogap, for simplicity we assume that this
quantity follows
a  linear
dependence on doping $W=W_{max}(1-x/x_1)$ ($x_1\simeq 0.19$),
while we let the critical temperature of a pure superconductor
$T_{c0}$ vanish at $x\simeq 0.3$, since in most experiments
superconductivity is not observed for doping value exceeding
$0.4\div 0.5$. It is clearly seen now how the superconducting area
shrinks while the scattering by normal impurities increases
($\Gamma$ is impurity scattering rate (34), normalized by
$T_{c0}$). This phase diagram is surprisingly similar to the
figure 5 of experimental paper \cite{Naqib}. } \label{phase}
\end{figure}

\section{Finite correlation length}

We now return to the problem of pseudo-gap formation in the
presence of order parameter fluctuations with finite correlation
length.  This problem was originally considered by Lee Rice and
Anderson\cite{LRA}. Following their approach, we consider Gaussian
fluctuations of a static order parameter with finite correlation
length and a real-space correlation function of the form
\begin{equation} \label{correlator}
<\Psi^{*}(x)\Psi(x')>=|\Psi|^2 exp(-\kappa{|x-x'|})e^{i 2p_F(x-x')},
\end{equation}
where $|\Psi|^2$ is the mean square fluctuation of the order
parameter field and $\kappa $ is the inverse correlation length.

Sadovskii \cite{Sadovskii74,Sadovskii79} later extended the FGM
model by considering the effect of multiple scattering off order
parameter fluctuations. The effect of superconducting pairing was
incorporated into the model in recent papers \cite{pairing}. In
the case of infinite correlation length, the Sadovskii yields an
essentially exact solution in one dimension. The Sadovskii
approach is only approximate at finite correlation length
\cite{Tchernysh}. For example, Bartosch and Kopietz
\cite{KopietzDOS} have shown, that the Dyson singularity in the
density of states, which exists for any finite value of $\xi$, is
missed by Sadovskii's algorithm.
Nevertheless, the Sadovskii approach seems
to be justified in 2D case for
certain topologies of the Fermi surface \cite{KuchSad}.
Nevertheless, the Sadovskii approach appears to provide a
remarkably good approximation to the density of states in a
pseudogap system with finite correlation length
\cite{Millis,SadphysicaC}.

To overcome some of the difficulties encountered in these early
treatments, Kopietz has recently suggested a simpler alternative
model, which exhibits the same static spatial correlations
originally considered by Lee Rice and Anderson, but which can be
treated exactly. Kopietz's approach considers an order parameter
of the form
\begin{equation}
\Psi(x) = A e^{i q  x}
\end{equation}
where the amplitude is a Gaussian random variable but the $q-$
vector is determined by a Lorentzian distribution, as follows
\begin{equation}
P(\bar A,A,q) = \frac{1}{2\pi^2 W ^{2}}e^{-\bar AA /2 W
^{2}}\frac{\kappa}{ (\kappa ^{2}+ (q-Q_{o})^{2})}
\end{equation}
where we have chosen a one dimensional
example, setting $Q_{o}=2p_{F}$.   The order parameter correlation function is
then
\begin{equation}\label{}
\langle \Psi^{*} (x) \Psi(x') \rangle = \int dA d\bar A dq P(\bar A, A , q)
\bar A A e^{i q(x-x')}= W^2 e^{-\kappa \vert x-x'\vert }e^{iQ_0(x-x')}.
\end{equation}
For the commensurate case this recovers the form (\ref{correlator})
first assumed by LRA.

The Hamiltonian which describes the coupling between the order
parameter fluctuations and the electrons is given by
\begin{equation}\label{kham}
  H(\bar A, A, q) = \sum_{k}\epsilon_{k}c\dg_k c_k +
\sum _{k} [Ac\dg _{k+q/2} c_{k-q/2}+ \bar A c\dg _{k-q/2}c_{k+q/2}].
\end{equation}
Unlike the scattering described in (9), here we have 
a quenched distribution of critical scattering $q-$vectors, and it is this
feature that gives rise to the finite correlation length.
By resuming all diagrams associated with scattering off the
fluctuating  field, Kopietz was able to derive the spectral
density for this model.

We now show how the methods we have developed in the previous
sections can be used to directly derive the Free energy for this
model, from which secondary properties, such as the spectral Green
function can also be derived.
The starting point for our discussion is the Free energy in a
given static configuration of the order parameter, given by
\begin{equation}\label{}
 F(\bar A,A,q)= -T \ln {\rm Tr} \bigl[e^{-\beta H(\bar A, A, q)}\bigr]
\end{equation}
Following our earlier discussion, the Gaussian fluctuations over
the static classical fields are now represented by the quenched
averaged Free energy
\begin{equation}\label{}
F=\int dA d\bar A dq P(\bar A, A , q) F(\bar A,A,q)
\end{equation}

The integrals over $A$, $\bar A$ and $q$ can be carried out
analytically (Appendix B) and yield the following result
\begin{equation} \label{}
F= -2T\int dx
 e^{-x}
{\rm  Tr}\ln [(\tilde{\omega }_{n})^{2} + (\epsilon _{k+ Q_{o}/2
})^{2} + 2x W^{2}  ],
\end{equation}
where $\tilde\omega_{n}=\omega_{n}+ (v_{F}\kappa /2) sign
\omega_{n}$. Thus the most important effect of the finite
correlation length is to introduce an {\sl imaginary} scattering
term inside the Greens function.  Notice that the distribution
over gap sizes has become exponential (Rayleigh), rather than
Gaussian, as it was in the case where $Q_{o}=\pi $ : this is a
consequence of having assumed an incommensurate $Q$ vector.  Thus
the essential physics of a finite correlation length manifests
itself as an elastic impurity scattering potential.

Various extensions of these arguments are possible, such as the
extension to two dimensions, the use of a commensurate scattering
potential and the introduction  of pairing terms, but we shall not
pursue them in this paper.

\section{Discussion and Conclusions}

The central idea of this paper, was to point out that a large
class of treatments of the pseudogap problem may be formulated in
terms of a quenched average over a distribution of static
classical order parameter fluctuations.   We have illustrated this
method, showing how it can be used to construct the Free energy of
the pseudo-gap system, from which mean-field equations can be
derived. The case of infinite correlation length is easier to
handle, but a simplified version of the case with finite
correlation length has also been treated.

The advantages of this method are that they permit us to begin
with the Free energy functional, rather than working directly with
Green functions.  Thus we are able to introduce superconducting
pairing into the formalism without having to reconsider a whole
class of two-particle correlation functions\cite{PosSad}.

Despite the utility of the new method, there are a number of
important questions that arise as to its complete validity. There
are two obvious weaknesses in the current approach

\begin{itemize}

\item The method ignores any frequency dependence in the
fluctuations.  For this reason, the method is only applicable at
high temperatures.

\item A secondary weakness is the assumption that the classical
modes are strictly non-interacting.  With this widely used
assumption, we have shown that the fluctuations behave as a source
of quenched disorder.  The equivalence between classical
fluctuations and quenched disorder only holds when the modes are
non-interacting and classical

\end{itemize}

Both weaknesses are clearly issues of time scales.  So long as the
characteristic time-scale of the electrons is longer than the
characteristic time scale of the fluctuations, then it is
reasonable to treat them in the way we have outlined.  If by
contrast, we are interested in  situations where the
characteristic electron time scales become very large, such as the
vicinity to a quantum critical point, then we might expect the
methods used here to become invalid.

Nevertheless, there are clearly a wide number of applications for
this approach.  One of the interesting possible applications are
transport properties of a pseudogap system, and these are
currently under active investigation.

{\bf Acknowledgments}  We would like  to thank J.O. Indekeu  for
early discussions  on this  work. Related  discussions with P.
Chandra , T. Giamarchi, E. Kuchinskii and especially M.V.
Sadovskii are also gratefully acknowledged. This work is
supported in  part by the National Science  Foundation grant
NSF-DMR 9983156  (PC). The hospitality of LVSM,  KULeuven and the
MPICPS, Dresden, Germany where parts of this work were carried
out, is highly appreciated.

\section{Appendix A}\label{}

The purpose of this appendix is to carry out the various
integrations inside the gap equation
\begin{equation}\label{a1}
\frac{1}{g}= 2T_{c} \int_{-\infty }^{\infty}d\zeta P (\zeta )
\left\{ \sum_{n, \ \kappa\in \frac{1}{2}\hbox{\tiny BZ}}
\frac{\gamma_{\kappa}^{2}}
{\omega_n^2+\epsilon_{\kappa}^2+\zeta^{2}W_{\kappa-Q/2} ^{2}}
 \right\}
.
\end{equation}
where we have denoted
\[
P (\zeta )= \frac{1}{\sqrt{2\pi }}e^{-\zeta ^{2}/2},
\]
If the pairing is dominated by processes near the Fermi energy, we
may replace the momentum sum by an energy integral, thus
\begin{equation}\label{a2}
\sum_{\vec{ k}} \rightarrow N (0)\int \frac{d\theta }{2\pi }
\int_{-\infty }^{\infty } d\epsilon
\end{equation}
so the gap equation becomes
\begin{equation}\label{a3}
\frac{1}{g}=2\pi N (0)T_{c}\int_{-\infty }^{\infty}d\zeta P (\zeta
) \int \frac{d\theta }{2\pi }\gamma ( {\theta })^{2} \left\{
\sum_{n=0}^{N_{max}-1} \frac{1 } {
\sqrt{\tilde{\omega}_n^2+\zeta^{2}W (\theta) ^{2}}}
 \right\}
.
\end{equation}
where we have introduced an upper cutoff $N_{max}= D/2\pi T_{c}$
into the Matsubara sum. To convert the Matsubara sum into a
contour integral, we need to identify the function $G (z)$that has
poles at $z=i\tilde{\omega }_{n}$, ($n\in \{0,N_{max}-1 \}$). By
using the identity
\[
\psi (z) = - C + \sum_{n=0}^{\infty}
\left(\frac{1}{1+n}-\frac{1}{z+n} \right),
\]
we see that
\[
\sum_{n=0}^{N_{max}-1} \frac{1}{z+n} = \psi (z+N_{max})-\psi (z) ,
\]
so that
\begin{equation}\label{a3x}
G (z)=-2\pi iT \sum_{n=0}^{N_{max}} \frac{1}{z-i\tilde{\omega}
_{n}} = \psi \left(\frac{1}{2}+ \frac{D+\Gamma +iz}{2\pi T}
\right)
-
\psi \left(\frac{1}{2}+ \frac{\Gamma +iz}{2\pi T}  \right)
\end{equation}
has the required properties. With this result, we can rewrite the
Matsubara sum as
\begin{eqnarray}\label{a4}
2\pi T_{c}\sum_{n=0}^{N_{max}} \frac{1 } {
\sqrt{\tilde{\omega}_n^2+\zeta^{2}W_{\theta }^{2} }}&=& 2\pi T_{c}
\int\frac{dz}{2\pi i } \left(\frac{i}{\sqrt{z^{2}-\zeta^{2}
W_{\theta }^{2}}} \right) [i G (z)] \cr
 &=& \int\frac{dz}{2\pi  }\frac{i}{\sqrt{z^{2}-\zeta^{2}
W_{\theta }^{2} }}G ( z)
\end{eqnarray}
where the integral is taken anticlockwise around the positive
imaginary axis.  
\fg{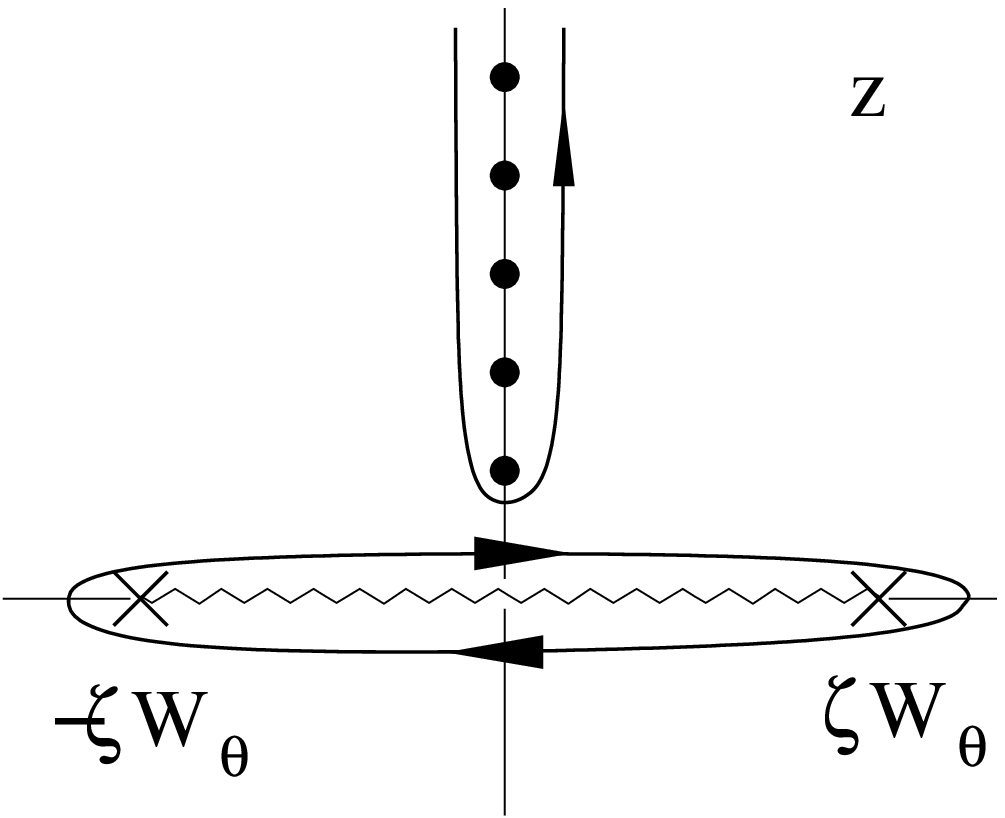}{fig7}{Showing how contour around the poles of $G (z)$ on
the positive imaginary axis in equation (49) is distorted around the
branch cut running between $\pm \zeta W_{\theta}$.}
Notice that we have analytically extended
$i\omega _{n}\rightarrow z$ by replacing  $
\sqrt{\tilde{\omega}_n^2+\zeta^{2}W_{\theta }^{2} } \rightarrow -i
\sqrt{z^{2}-\zeta^{2}W_{\theta }^{2} } $. With this choice, there
is a branch-cut running from $z= -\zeta W_{\theta }$ to $z=\zeta
W_{\theta }$. Above and below the branch-cut, $\sqrt{(x\pm i\delta
)^{2}-\zeta ^{2}W_{\theta }^{2}}
=
 \pm i \sqrt{\zeta ^{2}W_{\theta }^{2}-x^{2}}
$.  We may now distort the contour so that it runs clockwise
around this branch-cut, to obtain
\begin{equation}\label{a5}
2\pi T\sum_{n=0}^{N_{max}} \frac{1} {
\sqrt{\tilde{\omega}_n^2+\zeta^{2}W ^{2}}} = \int_{-\zeta
W_{\theta }}^{\zeta W_{\theta }}\frac{dx}{\pi
}\frac{1}{\sqrt{\zeta^{2}W_{\theta }^{2}-x^{2}} }G[x].
\end{equation}
Along the branch-cut, we may replace
\[
\psi \left(\frac{1}{2}+ \frac{D+\Gamma +iz}{2\pi T}
\right)\rightarrow \ln \left(\frac{D}{2\pi T} \right),
\]
so that inside the integral we can use
\begin{equation}\label{a7}
G[x]= \ln \left(\frac{D}{2\pi T} \right) - \psi \left(\frac{1}{2}+
\frac{\Gamma +ix}{2\pi T}  \right)
\end{equation}
At this point, the gap equation takes the form
\begin{equation}\label{a8}
\frac{1}{gN (0)}=\int \frac{d\theta }{2\pi } \gamma ( {\theta
})^{2} \int_{-\infty }^{\infty} d\zeta P (\zeta ) \left\{
\int_{-\zeta W_{\theta }}^{\zeta W_{\theta }} \frac{dx}{\pi}
\frac{1}{ \sqrt{ \zeta^{2}W_{\theta }^{2}-x^{2} } } G[x]
 \right\}
.
\end{equation}
We can further simplify this integral by reversing the order of
innermost $\zeta $ and $x$ integrations.  We do this as follows
\begin{eqnarray}\label{a9}
I=\int_{-\infty}^{\infty } \frac{d\zeta }{\sqrt{2\pi }} e^{-\zeta
^{2}/2} \int_{-\zeta W_{\theta }}^{\zeta W_{\theta }}
\frac{dx}{\pi} \left[\dots  \right] =\sqrt{\frac{2}{\pi ^{3}}}
\int_{-\infty}^{\infty } dx \int_{ \frac{|x|}{W_{\theta }
}}^{\infty }d\zeta e^{-\zeta ^{2}/2} \left[\dots  \right]
\end{eqnarray}
Next we make the change of variable, $u= \frac{\zeta
^{2}}{2}-\frac{x^{2}}{2W_{\theta }^{2}}$, so that the inner
integral becomes
\begin{equation}\label{a10}
I= \sqrt{\frac{2}{\pi ^{3}}}\int_{-\infty }^{\infty }dx
e^{-x^{2}/2W_{\theta }^{2}}\int_{0}^{\infty }du
\frac{e^{-u}}{\sqrt{2 (u+x^{2}/2W_{\theta }^{2})}}\left[\dots
\right]
\end{equation}
Replacing the argument inside the integral by
\[
\left[\dots  \right]= \frac{G[x]}{\sqrt{ \zeta^{2}W_{\theta
}^{2}-x^{2} }}= \frac{G[x]}{W_{\theta }\sqrt{2u}}
\]
we obtain
\begin{eqnarray}\label{a11}
I&=& \frac{1}{\sqrt{2\pi ^{3}}}\int_{-\infty }^{\infty
}\frac{dx}{W_{\theta } } e^{-x^{2}/2W_{\theta
}^{2}}G[x]\int_{0}^{\infty }du \frac{e^{-u}}{\sqrt{u
(u+x^{2}/2W_{\theta }^{2})}}\cr &=&\frac{1}{\sqrt{2\pi
^{3}}}\int_{-\infty }^{\infty }dx e^{-x^{2}/2}G ( x W_{\theta })
\int_{0}^{\infty }du \frac{e^{-u}}{\sqrt{u (u+x^{2}/2)}}
\end{eqnarray}
The integral on the right hand side can be carried out
analytically,
\begin{equation}\label{a12}
\int_{0}^{\infty }du \frac{e^{-u}}{\sqrt{u (u+t)}}= e^{t/2}K_{0}
(\frac{t}{2})
\end{equation}
where $K_{0} (x)$ is the modified Bessel function ($y=K_{0} (x)$
is the solution to the differential equation $x^{2}y´´ +x y'-
x^{2}y=0 $), so that
\begin{eqnarray}\label{a13}
I= \int_{-\infty }^{\infty }dx \Phi (x) G ( x W_{\theta })
\end{eqnarray}
where
\[
\Phi (x ) = \frac{1}{\sqrt{2\pi ^{3}}}e^{-x^{2}/4}K_{0} (x^{2}/4)
\]
is a normalized distribution function ($\int_{-\infty }^{\infty
}dx \Phi (x)=1$). Using this to replace the inner two integrals of
the gap equation (\ref{a8} ) we obtain
\begin{eqnarray}\label{a14}
\frac{1}{gN (0)} &=& \int \frac{d\theta }{2\pi } \gamma ( {\theta
})^{2} \int_{-\infty }^{\infty} dx \Phi  (x) \left[\ln
\left(\frac{D}{2\pi T} \right) - \psi \left(\frac{1}{2}+
\frac{\Gamma +i W_{\theta }x}{2\pi T}  \right) \right] \cr &=& \ln
\left(\frac{D}{2\pi T_{c}} \right)  - \int \frac{d\theta }{2\pi }
\gamma ( {\theta })^{2} \int_{-\infty }^{\infty} dx \Phi  (x) \psi
\left(\frac{1}{2}+ \frac{\Gamma +i W_{\theta }x}{2\pi T}  \right)
\end{eqnarray}
where we have used the normalization of the distribution function
and $\gamma (\theta )^{2}$ to extract the logarithm from the
integral. If we set $W=0$ and $\Gamma =0$ in this expression, we
obtain the familiar result
\[
\frac{1}{gN (0)}= \ln \left(\frac{D}{2\pi T_{c0}} \right)  - \psi
(\frac{1}{2})
\]
enabling us to eliminate the coupling constant from the gap
equation, writing
\begin{eqnarray}\label{a13}
\ln \left(\frac{T_{c}}{T_{c0}} \right) = \int \frac{d\theta }{2\pi
} \gamma ( {\theta })^{2} \int_{-\infty }^{\infty} dx \Phi  (x)
\left[\psi \left(\frac{1}{2} \right)- \psi \left(\frac{1}{2}+
\frac{\Gamma +i W_{\theta }x}{2\pi T}  \right) \right]
\end{eqnarray}

\section{Appendix B}\label{appb}

The purpose of this section is to average the Free energy over the
classical fields of the pseudogap.  If we write the electron field
in a two-component notation as
\[
C_{k}=
\begin{pmatrix}
c_{k+q/2}\cr c_{k-q/2}
\end{pmatrix},
\]
then in the presence of the scattering potential
\[
H_{I} = \sum _{k} [Ac\dg _{k+q/2} c_{k-q/2}+ \bar A c\dg _{k-q/2}c_{k+q/2}]
\]
the corresponding electron Greens function is given by
\begin{eqnarray}\label{gfn}
-\langle C_{k\alpha } (\tau )C\dg _{k\beta } (0)\rangle &=& {\cal
G}_{\alpha \beta } (k,\tau )\cr
{\cal G}_{\alpha \beta } (k,\tau )&=&T\sum_{n} {\cal  G} (k,i\omega
_{n})e^{-i\omega _{n}\tau }
\end{eqnarray}
where
\[
{\cal G} (\vec{ k},i\omega _{n})= [i\omega _{n}- \epsilon _{k+ (q/2)
\tau _{3}} - A \tau _{-}-\bar  A\tau _{+} ]^{-1}
\]
is the inverse propagator of the electron in the pseudogap field.
To obtain this expression we have kept scattering between states with
$k\sim \pm k_{F}$, neglecting scattering into high energy states with
$k\sim \pm ( k_{F}+n q)$.
The free energy is then given by
\begin{equation}\label{}
F=\int \frac{dAd\bar A}{2\pi \Delta ^{2}}e^{-\bar AA /2 W^{2}}
\int
dq\frac{1}{\pi (\kappa ^{2}+ (q-Q_{o})^{2})}
F[q,\bar A,A]
\end{equation}
where
\begin{eqnarray}\label{}
F[q,\bar A,A]&=& - T {\rm  Tr}\ln [-{\cal  G}^{-1} (k,i\omega _{n})]
\end{eqnarray}

To carry out the Lorentzian integral over q, we  first
linearize  the electron kinetic energy around the
wavevector $q=Q_{o}$, writing
\begin{eqnarray}\label{}
&&- T \int \frac{dq}{\pi }\frac{\kappa }{(q-Q_{0})^{2}+\kappa ^{2}}
{\rm  Tr}\ln
[i\omega _{n}- \epsilon _{k+ (Q_{o}/2)
\tau _{3}} - v_{F} (q-Q_{0}) - A \tau _{-}-\bar  A\tau _{+} ]\cr
&=&
- T \int \frac{dq}{\pi }\frac{\kappa }{q^{2}+\kappa ^{2}}
{\rm  Tr}\ln [i\omega _{n}- \epsilon _{k+ (Q_{o}/2)
\tau _{3}} - v_{F} q - A \tau _{-}-\bar  A\tau _{+} ]
\end{eqnarray}
This Lorentzian momentum integral has two poles in the complex plane
at $q-Q_{o}= \pm i \kappa $.  For $\omega _{n}>0$, the poles of the logarithm
are
in the upper half complex plane, so we complete the contour in the
lower half plane, picking up the sole contribution to the pole from
$q=-i\kappa $.  For the opposite sign, i.e.  $\omega _{n}<0$ we
complete the contour in the upper half plane. The result of this
procedure is
\begin{eqnarray}\label{}
F&=& -T\int \frac{dAd\bar A}{2\pi \Delta ^{2}}e^{-\bar AA /2
\Delta^{2}}
{\rm  Tr}\ln [i\tilde{\omega }_{n} - \epsilon _{k+ (Q_{o}/2)
\tau _{3}}  - A \tau _{-}-\bar  A\tau _{+} ],\cr
\tilde{\omega } _{n}&=& \omega _{n} + ( v_{F}\kappa/2)
{\rm sign} (\omega _{n})
\end{eqnarray}
showing that the effect of the finite correlation length is to
introduce a damping term into the propagator.  We can further simplify
this result by writing the the nesting condition
$\epsilon_{k+Q_{o}/2} =
-\epsilon_{k-Q_{o}/2} $
as
$\epsilon_{k+Q_{o}/2 \tau _{3}} =
\epsilon_{k-Q_{o}/2}\tau _{3} $, so that
\begin{eqnarray}\label{}
F&=& -T\int \frac{dAd\bar A}{2\pi \Delta ^{2}}e^{-\bar AA /2 \Delta^{2}}
{\rm  Tr}\ln [i\tilde{\omega }_{n} - \epsilon _{k+ Q_{o}/2}
\tau _{3}  - A \tau _{-}-\bar  A\tau _{+} ],\cr
&=& -2T\int \frac{dAd\bar A}{2\pi \Delta ^{2}}e^{-\bar AA /2 \Delta^{2}}
{\rm  Tr}\ln [(\tilde{\omega }_{n})^{2} + (\epsilon _{k+ Q_{o}/2
})^{2} + \bar A A  ],
\end{eqnarray}

To simplify the Gaussian integral over the pseudo gap amplitude, we simply
make the change of variables
\begin{eqnarray}\label{}
A&=& \sqrt{2x}We^{i\theta }\cr
\bar  A&=& \sqrt{2x}We^{-i\theta }
\end{eqnarray}
so that
\[
\frac{dAd\bar A}{2\pi W^{2}}e^{-\bar AA/2W^{2}}=
\frac{dxd\theta
}{2\pi } e^{-x}
\]
so that the Free energy becomes
\begin{eqnarray}\label{}
F= -2T\int dx
 e^{-x}
{\rm  Tr}\ln [(\tilde{\omega }_{n})^{2} + (\epsilon _{k+ Q_{o}/2
})^{2} + 2x W^{2}  ],
\end{eqnarray}
introducing
a Rayleigh
distribution of gap sizes.


\end{document}